\documentclass{article}

\usepackage{amssymb}
\usepackage[cp1251]{inputenc}
\usepackage[english]{babel}
\usepackage{amsmath}
\usepackage{color}
\usepackage{graphics}
\usepackage{epsfig}

\def\blf{$}
\def\elf{$}

\def\bdf{$$}
\def\edf{$$}

\def\beq#1{\begin{equation}\label{#1}}
\def\eeq{\end{equation}}

\title{Maslov's concept of phase transition from Bose-Einstein to Fermi-Dirac distribution. Results of interdisciplinary workshop in MSU.}

\author{A.S.Mishchenko \\ (Lomonosov Moscow State University)}
\date{15 October 2018}

\begin{document}
\maketitle

\begin{center}
The lecture delivered at first Chinese-Russian scientific and practical forum\\
"High technology: from science to implementation"

14-18 October 2018, Harbin, China.
\end{center}

\section{Introduction}
At the end of 2017, an interdisciplinary scientific seminar was organized at Moscow University, devoted to the study and development of a new scientific concept created by V.P. Maslov, allowing you to take a fresh look at the statistics of Bose-Einstein and Fermi-Dirac ideal gases. This new point of view allows us to interpret the indicated statistics as particular cases of statistical properties in number theory, on the one hand, and to indicate the limits of phase transitions from Bose to Fermi distributions.

Such a sort of phase transitions have already been observed in experiments on the decay of helium $ He_4$, which forms a Bose gas, into helium $ He_3$ and a fermion. The concept built by Maslov allows practical calculations of thermodynamic potentials of
Bose and Fermi gases in the region of phase transitions.

Scientists from Moscow State University and other Moscow scientific centers take part in the seminar: academician V.P. Maslov, academician A.T. Fomenko, corresponding member RAS V.E. Nazaikinsky,
the head of the department of general topology, Professor Yu.V.Sadovnichy, Professor of the RUDN A.Yu.Savin, Professor A.G.Kushner, and many of their students and employees. The result of the seminar was the development of a mathematical model for the calculation of thermodynamic potentials and their comparisons for Bose and Fermi gases.

\section{Formulation of the problem}

The Bose–Einstein formulas for the equilibrium occupation numbers in a Bose gas of noninteracting particles are well known in physics and gives the analogue in the number theory. Physicists claim that these formulas, which follow from the indistinguishability of particles, are only true in the quantum case, while in the classical case one should use the Boltzmann distribution.

However, V.P.Maslov put forward the idea that the same indistinguishability argument can be used to justify the Bose–Einstein distribution in the classical case, which, in conjunction with the
notion of fractional number of degrees of freedom, permitted him in particular to show how the van der Waals equation emerges for a classical gas of noninteracting particles.

He gave the first rigorous estimates for the probabilities of large deviations from
the Bose–Einstein distribution for large energies.

The Bose–Einstein distribution is closely related to the theory of partitions and number theory; for example, the case of two degrees of freedom corresponds to the partitio numerorum problem in number theory.

Let
$$
\lambda_1\geq\lambda_2\geq\lambda_3\geq\cdots\lambda_n\geq\cdots
$$
be a nondecreasing sequence of positive integers tending to \blf +\infty\elf.
 Given a positive integer \blf M\elf , consider
the Diophantine equation
$$
\sum\limits^{\infty}_{k=1}\lambda_k n_k  M
$$
with unknowns $ n_k\in \mathbb{Z}_{+} , k = 1,2,\dots$.

The Diophantine equation can be viewed as an
(integer-valued) model of a physical system of Bose particles (the number of particles is not fixed) in
which, for each \blf k\elf, \blf n_k\elf particles sit in the \blf k\elf th eigenstate (with energy \blf\lambda_k\elf ), the total energy of the system
being equal to \blf M\elf. The numbers \blf n_k\elf are known as the \blf occupation \, numbers\elf.

Note that the problem can be stated in a different way. For each \blf j\in\mathbb{N}\elf, let
\bdf
N_j=\sum\limits_{k:\lambda_k=j}n_k
\edf
be the total number of particles at the energy level \blf j\elf. Then the Diophantine equation becomes
\beq{}
\sum\limits_{j=1}^{\infty}jN_j=M, \quad \sum\limits_{j}N_j=N.
\eeq

It is a classical problem of partitioning of a number.

This problem can be formulated as: How many ways can a positive integer \blf M\elf be represented as a sum of
\blf N\leq M\elf positive integers, \bdf M = a_1 + a_2 + \cdots+ a_N ?\edf This problem statement is equivalent to: What is the
number of sets of integers \blf N_i\geq 0\elf such that equalities (1) are satisfied?
It is obvious that any set of integers \blf N_i = 0\elf satisfying system (1) corresponds to
one partition of \blf M\elf.

We must note that in the book by Landau and Lifshitz and in
quantum thermodynamics, where bosons and fermions are considered, two equations are used: one of them
contains the number \blf N\elf of particles, and the other contains the energy \blf E\elf. The number of particles can be
regarded as an integer, but the energy \blf E\elf is not an integer (it even has a dimension)

The fundamental number theory problem of decomposing a number \blf M\elf into \blf N\elf terms in the space of
integers splits into two cases: repeated terms are admissible in the first case and not admissible in the second
case.

It is well known that the first case (with repeated terms) corresponds to the Bose–Einstein distribution, and the second case (without repeated terms) corresponds to the Fermi–Dirac distribution. This fact
indicates that analytic number theory must be related to statistical physics. Here, we solve the analytic
continuation problem in the case of a transition from Bose statistics to Fermi statistics.

We let \blf N_i\elf denote the number of particles at the \blf i\elf th energy level. In the case of Gentile statistics,
there can be at most \blf K\elf  particles at each energy level. According to the generally accepted definition,
we have \blf K = 1\elf for the Fermi–Dirac distribution and \blf K =\infty\elf for the Bose–Einstein distribution. But it
obviously follows  that \blf N_i \leq N\elf for the Bose system, and hence \blf K\leq N\elf for the Bose system. It
hence follows that the maximum \blf K \elf is equal to \blf N\elf and not infinity. This physical conclusion substantially
changes the formulas.

The mass \blf m\elf and the Planck constant \blf \hbar\elf were introduced as parameters in Landau and Lifshitz book. In number theory, we
can assume that all these constants and the volume \blf V\elf are equal to unity. To compare number theory
formulas and statistical distributions, we introduce the notation for some quantities that are equal to unity
in number theory.

 Let us introduce a general parameter \blf \Phi\elf that is consistent with the notation in the book by Landau and
Lifshitz :
\bdf
\Phi=\lambda^{2(\gamma+1)}VT^{\gamma+1}=V(\lambda^{2}T)^{\gamma+1},
\edf
where \blf V\elf is the volume, \blf T\elf is the temperature, and
\blf \lambda=\sqrt{2\pi m}/2\pi\hbar\elf.
\vskip 1cm

Then in the case \blf N = K\elf, we  obtained crucial new equations for the energy \blf E\elf  and the number \blf N\elf  of particles:

\bdf
E=\Phi T(\gamma+1)\left(Li_{2+\gamma}(a)-\frac{1}{(N+1)^{\gamma+1}}Li_{2+\gamma}(a^{N+1})\right),
\edf

\bdf
N=\Phi \left(Li_{1+\gamma}(a)-\frac{1}{(N+1)^{\gamma}}Li_{1+\gamma}(a^{N+1})\right),
\edf
where \blf\gamma=D/2-1\elf, \blf D\elf is the number of degrees of freedom, and \blf a\elf is the activity. Here \blf Li_s(z)\elf is the polylogarithm function:
\bdf
Li_s(z)=\sum\limits_{k=1}^{\infty}\frac{z^k}{k^s}
\edf

\section{Results}
In the case of Gentile statistics, the self-consistent equations  have the form
\bdf
N=\Phi T(\gamma+1)\left(Li_{2+\gamma}(a)-\frac{1}{(N^\alpha+1)^{\gamma+1}}Li_{2+\gamma}(a^{N^\alpha+1})\right),
\edf

\bdf
M=\Phi \left(Li_{1+\gamma}(a)-\frac{1}{(N^\alpha+1)^{\gamma}}Li_{1+\gamma}(a^{N^\alpha+1})\right).
\edf

The value \blf \alpha\elf varies from \blf 0\elf to \blf 1\elf. For \blf \alpha = 1\elf , the Bose distribution occurs, while for \blf \alpha = 0\elf , the Fermi
distribution occurs. When \blf \alpha = 0.5\elf , the Fermi distribution for the main term of the asymptotics
occurs. In the interval between \blf \alpha = 0\elf  and \blf \alpha = 0.5\elf , there is a concentration of curves depicted in
Figures 3 and 4.

\includegraphics[width=0.5\textwidth]{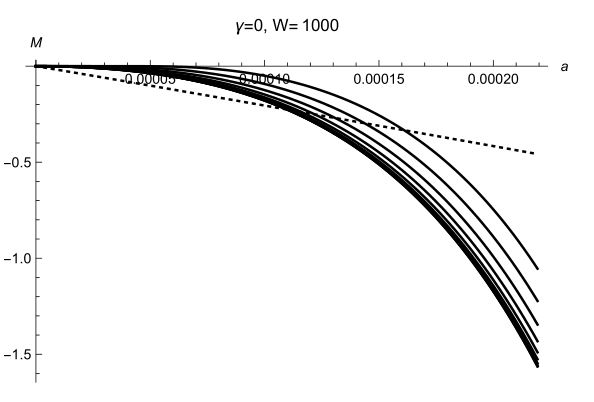}

Fig. 3. Dependence of \blf M\elf on a obtained numerically. The solid curves correspond to \blf W = 1000,
 \gamma= 0, \alpha = 0.9, 0.8, 0.7, 0.6, 0.5, 0.4, 0.3, 0.2, 0.1,0.01\elf (from right to left). The dotted line
is \blf N = -1/log(a)\elf.

\includegraphics[width=0.5\textwidth]{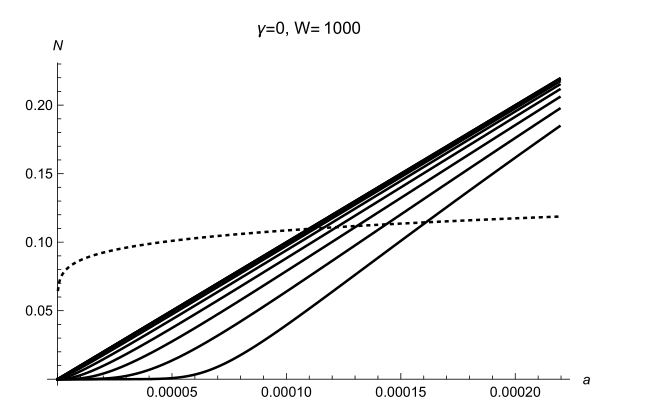}

Fig. 4. Dependence of \blf N\elf on \blf a\elf obtained numerically  corresponding to the plot of \blf N(a)\elf turned
over. The solid curves correspond to \blf W = 1000,
 \gamma= 0, \alpha = 0.9, 0.8, 0.7, 0.6, 0.5, 0.4, 0.3, 0.2, 0.1,0.01\elf (from right to left). The dotted line
is \blf N = -1/log(a)\elf.

So it was shown that, depending on the hidden
parameter, purely quantum problems behave like classical ones.  It
is shown that the Bose-Einstein and the Fermi-Dirac distributions, which until now were
regarded as dealing with quantum particles, describe, for the appropriate values of the hidden
parameter, the macroscopic thermodynamics of classical molecules.

We stress that the transition of a boson into a fermion occurs not at the point \blf a = 0\elf but at a nonzero
\blf a\elf. We let \blf a_0\elf denote a value of the activity \blf a\elf at which the number of particles in the Bose-Einstein
distribution is \blf N = 0\elf. At this point, there are already no bosons. This is the instant of transition when
the boson has already disappeared, splitting into two fermions, and one of the fermions must somehow
be lost. What happens with it? At what distance from the nucleus shell does it move? This can be
calculated by determining the energy value required for boson decay. This energy is very small although
the fermion separated in a macroscopic volume \blf V\elf .

When one of the fermions leaves this volume, this means
that it disappears. Namely, this means that the fermion separates not in the nucleus shell but in a greater
volume. In this situation, there is a jump of spin in a macroscopic volume. This conclusion agrees well
with experiments, in particular, described by Bell.

We note that the fermions do not interchange, which
means that the fermions are numbered. This energy of the spin jump as the specific energy can be
calculated.

The jump of the compressibility factor from the Fermi system into the Bose system is determined by
the formula
\bdf
\Delta Z(a)=Z|_{Fermi}-Z|_{Bose}=
\frac{Li_{2+\gamma}(-a)}{Li_{1+\gamma}(-a)}-
\frac{Li_{2+\gamma}(a)}{Li_{1+\gamma}(a)}.
\edf


\vskip 1cm


\begin{thebibliography}{aaa}
\bibitem{bell}
 J. S. Bell, “On the Einstein Podolsky Rosen paradox,” Phys., 1, No. 3, 195–200 (1964).

\bibitem{Maslov-2018}
{\sc V.P.Maslov,}
{\it New Formulas related to Analytic Number Theory
and their Applications in Statistical Physics,}
{\rm Theoretical and Mathematical Physics, 196(1): 1082–1087 (2018)}




\bibitem{Maslov-Mishchenko-2003}
{\sc V.P.Maslov and A.S.Mishchenko,}
{\it Geometry of a Lagrangian Manifold in Thermodynamics
(Principle of Minimizing the Thermodynamical Potential and Thermodynamical
Inequalities.
Analysis of the Gibbs Method of Geometric Picture of Thermodynamics),}
{\rm Russian Journal of Mathematical Physics}
{\rm  Maik Nauka/Interperiodica Publishing (Russian Federation),2003, tom 10, No. 2, p. 161-172 }

\bibitem{Maslov-Nazaikinskii-2016}
{\sc V.P.Maslov and V.E.Nazaikinskii,}
{\it On the Rate of Convergence to the Bose–Einstein Distribution}
{\rm Mathematical Notes, 2016, Vol. 99, No. 1, pp. 95–109}
\end{thebibliography}
\end{document}